\documentclass[epjc3,twocolumn]{svjour3}

\RequirePackage{graphicx}
\usepackage{subcaption}

\usepackage[english]{babel}
\usepackage{newtxtext,newtxmath}
\usepackage{amsmath,mathtools}
\usepackage[separate-uncertainty=true, multi-part-units=brackets]{siunitx}
\DeclareSIUnit{\slpm}{slpm}
\DeclareSIUnit{\ppb}{ppb}
\DeclareSIUnit{\PE}{PE}
\DeclareSIUnit{\tenns}{10ns}
\usepackage[version=4]{mhchem}
\usepackage{tikz}
\usepackage{tabularx}
\usepackage{comment}
\usepackage{upgreek}
\usepackage{xcolor}
\usepackage{booktabs}
\usepackage{layouts}
\usepackage{hyperref}   
\usepackage{orcidlink}

\journalname{Eur. Phys. J. C}

\begin{document}
\sloppy

\title{First demonstration of long-distance electron drift in liquid xenon with the Xenoscope time projection chamber}

\date{Received: date / Accepted: date}

\author{
N.~Angelides \orcidlink{0000-0003-3930-934X} \thanksref{addr1}
\and
M.~Babicz \orcidlink{0000-0002-1017-5440} \thanksref{addr1}
\and
E.~E.~Barillier \orcidlink{0000-0001-9107-6542} \thanksref{addr1}
\and
L.~Baudis \orcidlink{0000-0003-4710-1768} \thanksref{addr1}
\and
H.~Birch \orcidlink{0000-0001-7476-8045} \thanksref{addr1}
\and
C.~Capelli \orcidlink{0000-0003-3330-621X} \thanksref{addr1,addr2}
\and
J.~J.~Cuenca-Garc\'ia \orcidlink{0000-0002-3869-7398} \thanksref{addr1}
\and
S.~George \thanksref{addr1}
\and
R.~N.~Hampp \orcidlink{0009-0005-1354-7616} \thanksref{addr1,email1}
\and
M.~A.~Hernandez \orcidlink{0000-0001-6483-0052} \thanksref{addr1}
\and
L.~H\"otzsch \orcidlink{0000-0003-2572-477X} \thanksref{addr1}
\and
F.~J\"org \orcidlink{0000-0003-1719-3294} \thanksref{addr1}
\and
K.~Mor{\aa} \orcidlink{0000-0002-2011-1889} \thanksref{addr1}
\and
M.~Noia \orcidlink{0009-0008-5529-8613} \thanksref{addr1}
\and
S.~Ouahada \orcidlink{0009-0007-4161-1907} \thanksref{addr1,email2}
\and
B.~Penning \orcidlink{0000-0001-9020-9112} \thanksref{addr1}
\and
F.~Piastra \orcidlink{0000-0001-8848-5089} \thanksref{addr1}
}

\institute{Physik-Institut, University of Zurich, Winterthurerstrasse 190, 8057 Zurich, Switzerland \label{addr1}
}

\thankstext{addr2}{Now at Department of Physics, Sapienza University of Rome and INFN Sezione di Roma, Piazzale Aldo Moro 2, 00185 Rome, Italy}
\thankstext{email1}{rebecca.hampp@uzh.ch}
\thankstext{email2}{sana.ouahada@uzh.ch}



\maketitle

\begin{abstract}

Xenoscope is a dual-phase xenon time projection chamber operated at the University of Zurich to investigate electron transport over distances relevant to the next-generation XLZD (XENON–LUX–ZEPLIN–DARWIN) observatory. We demonstrate electron drift over \SI{2.6}{\meter} in liquid xenon using cosmic-muon-induced ionization signals. External muon tagging provides a timing reference for charge produced at known depths, while the endpoint of the average electroluminescence waveform probes transport over the full drift length. We measure electron drift velocities of \SI{1.320 \pm 0.006}{\milli\meter\per\micro\second} at \SI{80.92\pm0.36}{\volt\per\centi\meter} and \SI{1.436 \pm 0.001}{\milli\meter\per\micro\second} at \SI{96.0 \pm 0.42}{\volt\per\centi\meter}. These results establish detectable charge transport over meter-scale distances and provide an experimental benchmark for the development of future large liquid-xenon time projection chambers.

\end{abstract}

\section{Introduction}
\label{sec:intro}

Dual-phase xenon time projection chambers (TPCs) are the most sensitive detector technology for weak-scale dark matter and other rare-event searches such as second-order weak processes \cite{Baudis_2023_TPC}. Experiments with multi-ton target masses and unprecedentedly low background levels \cite{XENON:2022ltv,XENON:2025nic,PandaX-4T:2021lbm,LZ:2022ysc} lead the exploration of the dark matter direct detection parameter space in the \SI{}{GeV}-\SI{}{TeV} mass range. Achieving reliable operation of a detector with a target mass of several tens of tonnes places high requirements on detector design and operation.

The planned XENON-LUX-ZEPLIN-DARWIN (XLZD) observatory will use a dual phase xenon TPC with an active target of 60--80\,t~\cite{Baudis:2024jnk,XLZD2025DesignBook}. XLZD will extend sensitivity to dark matter interactions down to the irreducible background from solar and atmospheric neutrinos~\cite{OHare:2021utq} (referred to in the literature as \emph{neutrino fog}) with exposures of approximately 100 tonne-years, while enabling searches for neutrinoless double-beta decay and other precision and rare-event measurements.

A dual-phase TPC operates by collecting prompt scintillation light (S1) and ionization electrons resulting from energy deposits in the liquid xenon (LXe). The electrons are drifted upwards using an electric field, towards the liquid–gas interface, where they are extracted by a stronger electric field and generate a delayed electroluminescence signal (S2) in the gas. The time separation between S1 and S2 provides the interaction depth, while the S2 light distribution determines the horizontal position. Combined, these signals allow for 3D event reconstruction and discrimination between electronic and nuclear recoils from their relative amplitudes. Consequently, all measurements rely critically on good knowledge of electron drift velocity, diffusion and survival probability, motivating characterization over the relevant scale for XLZD. 

Two complementary large scale test facilities using approximately \SI{400}{kg} of LXe have been developed to target different aspects of the XLZD scale-up program. The PANCAKE facility~\cite{Brown:2023vgf} focuses on constructing and running a large diameter detector, providing infrastructure for a TPC of up to \SI{2.6}{m} in diameter, whereas Xenoscope focuses on the challenges associated with long electron drift distances through a vertical TPC of \SI{2.6}{m} height. 

Beyond providing a platform for studying charge transport over unprecedented drift lengths, including measurements of electron attenuation and diffusion under controlled conditions, Xenoscope demonstrates stable operation of a meter-scale dual-phase detector. The facility will also enable dedicated investigations of the optical properties of LXe, such as Rayleigh scattering and vacuum-ultraviolet (VUV) light absorption by residual impurities, which are essential inputs for the design and optimization of next-generation detectors. Further applications include a feasibility study of coating-based radon barriers in a LXe detector\,\cite{Jorg:2026lse}.

In this paper, we present a first measurement of muon-induced electron signals drifting over the full length of the Xenoscope TPC.

\section{The Xenoscope facility}
\label{sec:facility}

The first phase of the Xenoscope experimental program used a \SI{53}{cm} purity monitor to characterize xenon purification and measure electron transport properties \cite{Baudis_2023}. This was followed by the installation of a \SI{2.6}{m} tall dual-phase TPC, whose construction and commissioning are reported in Ref.~\cite{Adrover_2025}. The design and construction of the facility are described in Ref.~\cite{Baudis_2021}.

The detector is housed in a double-walled stainless steel cryostat with a capacity of approximately \SI{400}{kg} of LXe. The cryogenic infrastructure, storage and recovery systems, together with the instrumentation for monitoring and controlling the facility, are described in Refs.~\cite{Baudis_2021,Baudis_2023}.

The TPC contains approximately \SI{150}{kg} of LXe within a cylindrical drift region with a nominal length of \SI{2.6}{m} and an internal diameter of \SI{16}{cm}. Its field cage consists of 173 oxygen-free high thermal conductivity copper rings supported by six polyamide-imide pillars and interconnected by a dual resistor chain. A stainless steel cathode plate defines the lower end of the drift region, while the gate and anode are located at the top. These two electrodes are hexagonal meshes welded into a \SI{4}{mm} wide circular frame. They are vertically separated by \SI{12}{mm}, with the liquid-gas interface between them. High voltage is delivered to the cathode through a feedthrough entering from below the cryostat. The electrode assembly, high-voltage delivery and level-control systems are described in Ref.~\cite{Adrover_2025}.

Light signals are detected by a silicon photomultiplier (SiPM) array containing 48 Hamamatsu S13371-6050CQ-02 square modules. The array is composed of 192 6$\times$6~mm$^2$ multi-pixel photon counters (MPPCs), distributed over 12 tiles~\cite{Peres_2023}. Each tile combines the signals from four square modules into one amplified analogue output. The photosensitive plane is positioned \SI{15}{mm} above the anode and faces the xenon volume. A blue LED coupled to optical fibers provides light for the calibration of the photosensor gains. The waveforms are recorded using two CAEN V1724 digitizers. Since inner detector surfaces are not covered with PTFE, light collection efficiency deteriorates rapidly along the length of the TPC.

The cathode plate at the bottom of the TPC is equipped with a photocathode equivalent to the one used in~\cite{Adrover_2025}. The photocathode is produced in-house by depositing a \SI{50}{\nm} thick gold film on a \SI{2}{\mm} thick fused-silica substrate disc following the procedure described in~\cite{Baudis_2023}. Light pulses from a xenon flash lamp are intended to liberate up to $10^6$ electrons via the photoelectric effect, mimicking a large S2 signal. However, no electrons emitted from the photocathode were observed despite regular flash lamp measurements. Further investigations are ongoing.

Two external BC-412 plastic scintillator panels of dimensions \SI{35.4}{cm} $\times$ \SI{5.7}{cm} $\times$ \SI{5.7}{cm} are placed on opposite sides of the outer cryostat vessel to tag muons that traverse the LXe. The panels are adjusted to select muons passing nearly horizontally. Each panel is read out by a three-inch PMT. The PMT signals serve two purposes: they are digitized together with the SiPM data, and they are fed into external coincidence electronics, which generates a trigger when both panels register a pulse well above the noise floor within \SI{40}{\ns}. The panels can be mounted at different heights outside the cryostat, which allows muons to be selected in different regions of the TPC. The coincidence trigger provides a timing reference for the passage of the muon, relative to which the arrival times of the S2 pulses recorded by the SiPM array are measured. The photosensor readout and muon-tagging instrumentation are detailed in Ref.~\cite{Adrover_2025}. A third panel placed above the TPC also provides an anti-coincidence veto for rejecting muon-induced showers.
\section{Data taking}
\label{sec:meas}

Prior to data taking, the LXe was purified for one month by continuously extracting and circulating gaseous xenon through a hot zirconium getter. The purification continued without interruption throughout the data taking period which began in late April and ended in late May of 2026.
The concentration of H$_2$O in the extracted GXe was measured using a commercial cavity ring-down spectrometer (HALO KA-M7500-S). It was found to be \SI{22.2\pm1.0}{\ppb}, closely matching the expected value based on the vapor pressure of water-ice at the LXe operating temperature\,\cite{Bsc_nauer}.

Data-taking alternated at regular intervals between xenon flash lamp, muon coincidence, and external trigger measurements, with photosensor calibrations performed in between. The muon coincidence selects events with electrons produced at a known depth, while the external trigger provides an independent reference for validation.

\subsection{SiPM calibration and stability}

\begin{figure}
    \centering
    \includegraphics[width=1\linewidth]{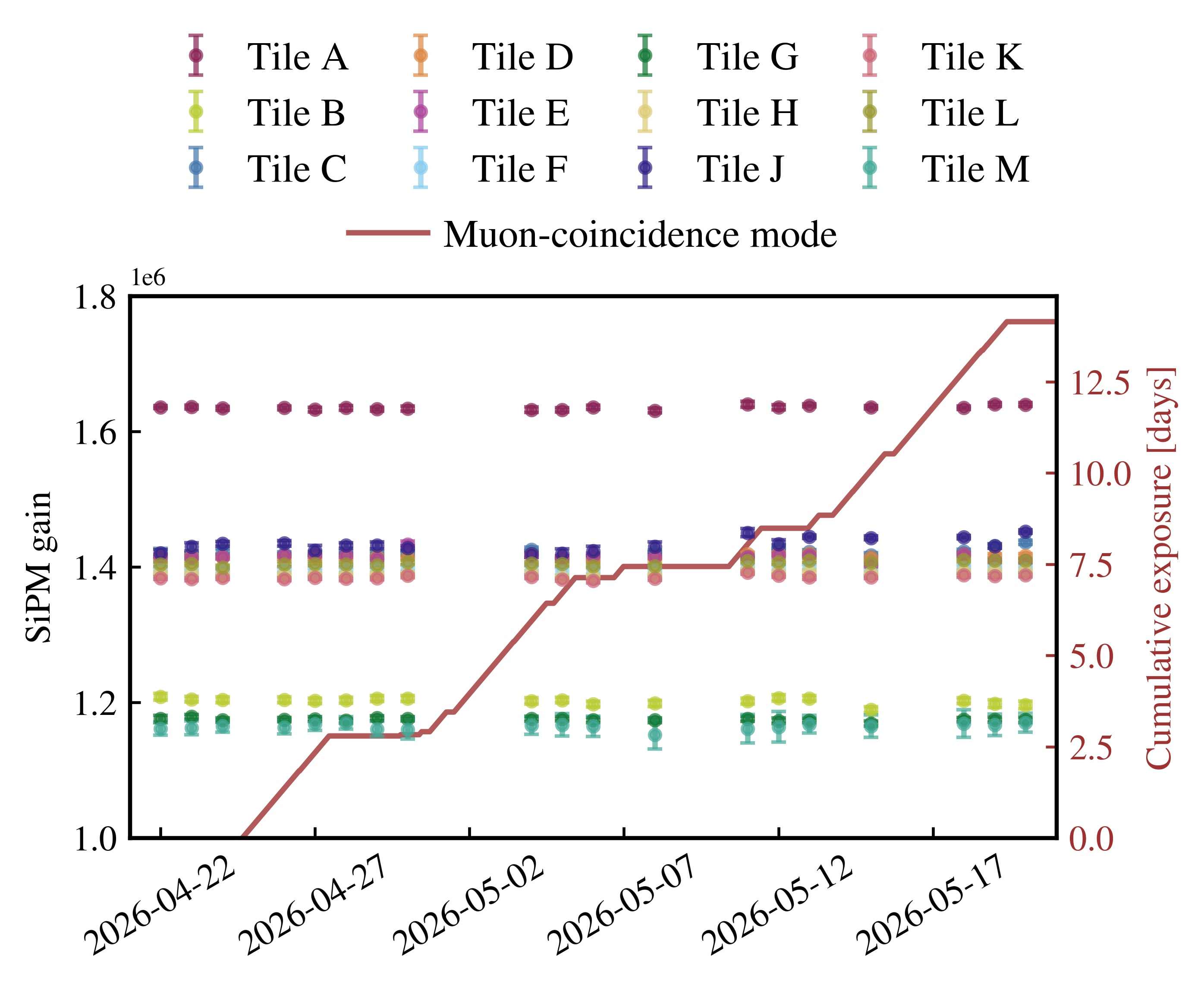}
    \caption{Stability of the SiPM tile gains throughout the run. The gain of each tile is extracted from a fit to the SPE charge spectra of daily LED calibrations taken at low light intensity. The cumulative exposure of regular muon measurement is shown in brown.}
    \label{fig:sipm_stability}
\end{figure}

The photosensors in the SiPM array were characterized prior to detector commissioning. For this run, we determine the optimal bias voltage of \SI{51}{\volt} and multiple LED illumination levels, to establish the single-photoelectron (SPE) response for all tiles. This response is then used both to monitor the SiPM gains throughout the run and to scale the signal of each individual tile before summing the waveforms. Details on the SiPM calibration can be found in App.~\ref{app:sipm}.

The resulting gain values are shown in Fig.~\ref{fig:sipm_stability}. The twelve tiles exhibit different absolute gain values, as also observed during the commissioning run, ranging approximately from $1.2\times10^6$ to $1.6\times10^6$. The gain of each individual tile remains stable throughout the run, with a relative standard deviation below 0.7$\%$ for all tiles. Most tiles show a relative SPE resolution between 6$\%$ and 8$\%$, with tile M standing out as an outlier at 11$\%$, suggesting differences in sensor response and readout performance.  

\subsection{Externally triggered measurements}

Regular externally triggered measurements are conducted during the run using a waveform generator to provide triggers at frequencies up to \SI{3}{\Hz} with the coincidence and anti-coincidence muon panels not recorded. This measurement is used as a background dataset to validate the event selection applied to the muon coincidence measurements. Regular short duration measurements yield a total cumulative exposure of \SI{4.85}{\hour}.

\subsection{Muon coincidence measurements}
\label{sec:mnt_des}

The two muon panels are mounted at an angle of $12.4^\circ$ relative to the horizon, with one positioned higher than the other, such that coincident muons cross the TPC along a well-defined diagonal path. The lower edges of the left and right panels are positioned \SI{6.4}{\cm} and \SI{37.4}{\cm} above the cathode and have a horizontal distance of \SI{141}{\cm}. This translates to a drift length of \SI{230.9}{\cm} to \SI{237.0}{\cm}. The minimum drift length corresponds to a muon grazing the top edges of both muon panels, while the maximum drift length arises from a muon grazing their bottom edges. We therefore assume that the mean muon interaction depth is normal distributed with mean \SI{233.95}{\centi\meter} and standard deviation of $6.1/\sqrt{12}=\SI{1.76}{\cm}$, assuming that the muon track interaction depth is uniformly distributed across the allowed range. The total exposure of the muon coincidence measurement after data quality selections is \SI{339.46}{\hour}. The cumulative exposure over time is shown in Fig.~\ref{fig:sipm_stability}.

\section{Analysis}
\label{sec:analysis}

This section describes the expected signal, reconstruction and selections applied to the data. Due to the lack of reflective PTFE in the TPC, S1 signals are not detected with sufficient efficiency for use in this analysis. The prompt scintillation signals are therefore not used in this analysis.

\subsection{Expected muon-induced signal}

Muons are produced in the atmosphere at a typical height of \SI{15}{\km} and are the most abundant charged particles at sea level. The integral intensity of muons above \SI{1}{\GeV} at an angle of $15^\circ$ above the horizon is \SI{4.743 e-4}{\per\square\centi\metre\per\steradian\per\second}~\cite{Jokisch:1979gh}. They are a convenient naturally occurring calibration source in a TPC located above ground.

The expected signal for this analysis is created by a muon crossing the TPC at a depth of \SI{233.95}{\cm} at the full diameter and an angle of 12.4$^\circ$. The muon creates a maximum ionization track length of \SI{15.3}{\cm} with a vertical distance of \SI{3.2}{\cm}. This corresponds to a pre-drift pulse width of \SI{23.6}{\us}, assuming a drift velocity of \SI{1.35}{\milli\metre\per\micro\second}~\cite{Jorg:2021hzu}. The longitudinal diffusion of the electron cloud over the mean drift length is approximately \SI{2.8}{\us}, rendering its contribution to the total pulse width subdominant\,\cite{Baudis_2023}. The linear stopping power of a \SI{10}{GeV} muon in LXe is \SI{~4.75}{\MeV\per\cm} assuming the Bethe-Bloch formula~\cite{Groom:2001kq}. Alongside an ionization length of \SI{15.33}{\cm}, this corresponds to $6.3 \times 10^6$ liberated electrons for a mean excitation energy of \SI{11.5}{\eV}~\cite{Baudis:2021dsq}. Assuming a single-electron gain of approximately 46 photoelectrons (PE) at a 20\% photon detection efficiency for gas scintillation, the resulting S2 signal from a \SI{10}{\GeV} muon is expected to have an integrated area on the order of $10^8$~PE and a width up to $\sim$\SI{27}{\us}. While this represents an idealized signal for a full diameter crossing at a 12.4$^\circ$ angle and mean drift length of \SI{233.95}{\centi\meter}, muons with grazing trajectories or smaller incidence angles will produce smaller signals.

\subsection{Reconstruction}

The raw waveforms are processed with a custom pulse-finding algorithm applied to the summed waveform detected by the array. First, a difference-of-Gaussians (DoG) filter, which subtracts the waveform smoothed with two Gaussian kernels of different widths and acts as a band-pass filter, is applied to enhance pulse features~\cite{2020SciPy-NMeth}. A candidate pulse lies between two consecutive minima of the filtered waveform. 
It is accepted as a reconstructed pulse only if the raw amplitude exceeds \SI{5}{\PE\per\tenns} at a local maximum of the filtered waveform, and if the mean raw amplitude over the candidate exceeds \SI{1}{\PE\per\tenns}. Pulses separated by less than \SI{1}{\micro\second} are merged into a single S2 peak. Each peak is assigned a drift time, defined as the time difference between the trigger time and the start of the peak, given by the first local minimum of the filtered waveform. The results were found to be robust against variations of the DoG smoothing scales, as tested on the resolution of the drift time peak. Further details of the reconstruction, including the smoothing scales and the stability of the method with respect to these choices, are given in App.~\ref{app:reconstruction}.

\subsection{Event selection}

Four selection criteria are applied to increase the signal significance while ensuring no artificial peak is introduced. Initially events without a reconstructed pulse are discarded. 
Following this, two selections are applied based on the coincidence and anti-coincidence panels. 
To reject triggers caused by accidental gamma-ray coincidences, events are accepted only if pulses are detected in both coincidence panels within \SI{500}{\nano\second} and their summed area exceeds \SI{1000}{ADC}. 
The anti-coincidence veto rejects events with a pulse area above \SI{140}{ADC} within the same coincidence window. Both selections are optimized on a dedicated muon measurement performed after the run and are discussed in detail in App.~\ref{app:muon_coin_tuning}.

The last selection exploits the pulse width, defined as the full width at half maximum (FWHM) and retains events in which the largest pulse has a width in the range of 5--\SI{30}{\micro\second}. This selection criterion is optimized on the drift time distribution of the main analysis dataset by maximizing the efficiency ratio defined as: the fraction of events retained in the drift time signal window relative to a background window. The signal window is defined to be 1600--\SI{1800}{\micro\second} based on the expected drift time assuming a drift velocity of \SI{1.35}{\milli\metre\per\micro\second}~\cite{Jorg:2021hzu}. The background window is chosen  between the trigger and the signal window at 400--\SI{1200}{\micro\second}. The resulting upper bound of \SI{30}{\micro\second} is consistent with the maximum expected pulse width of the signal. 

\begin{figure}
    \centering
    \includegraphics[width=\linewidth]{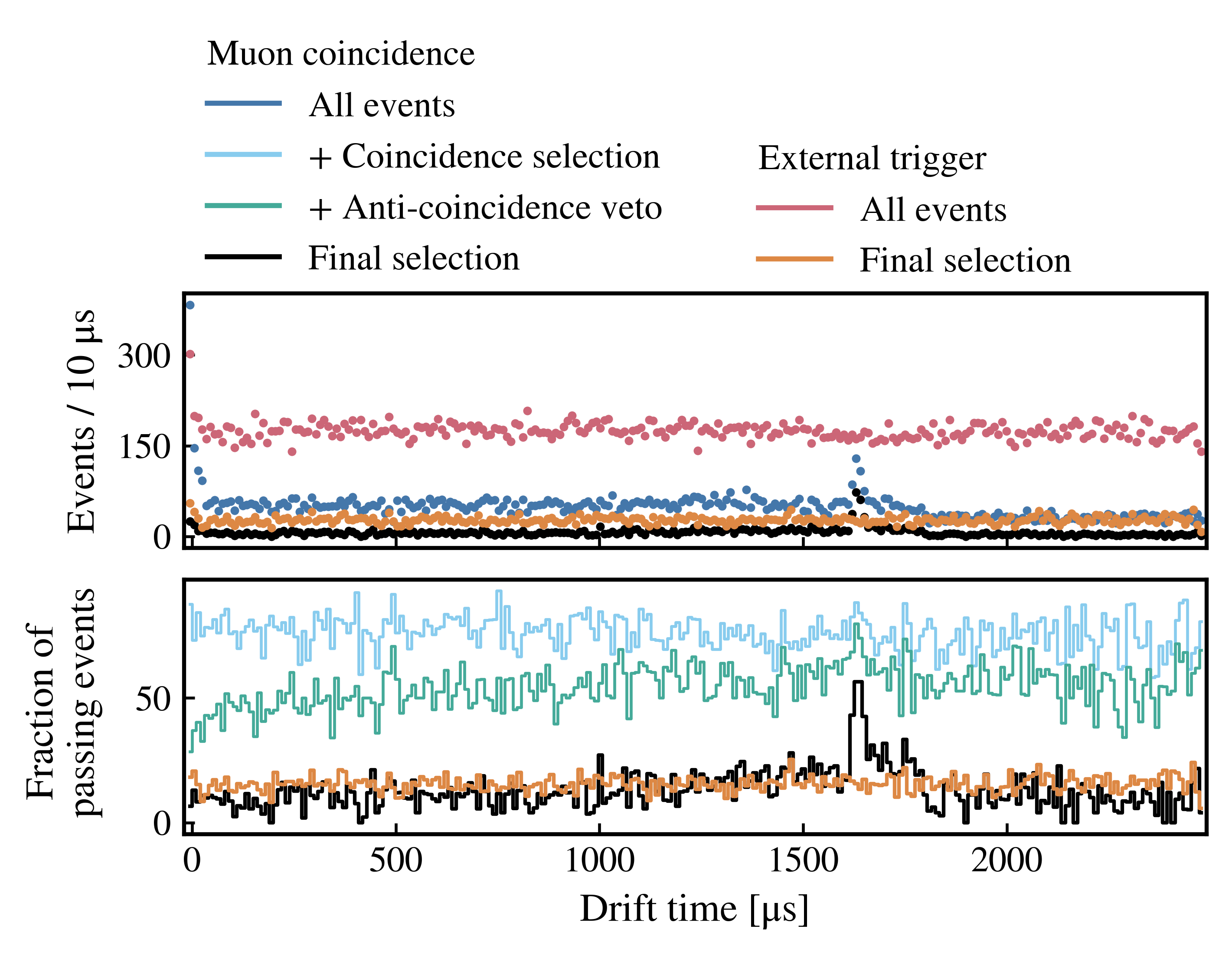}
    \caption{Top: Influence of the final selection on the pulse drift time distribution in the muon coincidence and external trigger dataset. Both datasets show a clear peak around zero due to vertical muon showers. A single negative bin is also visible, corresponding to reconstructed pulses that precede the trigger. Bottom: Fraction of events passing each selection step for both datasets. The selections clearly enhance the signal in the expected drift time interval. The acceptance of the externally triggered dataset is drift time independent.}
    \label{fig:WidthSelection}
\end{figure}

The pulse drift time distribution after each selection step is shown in Fig.~\ref{fig:WidthSelection} for both background and muon-coincidence data. The selections further enhance the drift time peak in the muon coincidence dataset, while the externally triggered data confirms that no selection introduces an artificial peak. The anti-coincidence veto successfully reduces the peak around zero drift time. The total number of events and the acceptance of each selection step are shown in Table \ref{tab:Run6_stats}.

\begin{table}[htbp]
    \centering
    \caption{Number of events and acceptance of the muon-coincidence measurements with an exposure of \SI{339.46}{h}.}
    \label{tab:Run6_stats}
    \begin{tabular}{l c c}
        \toprule
        \textbf{Selection step} & \textbf{Number of events} & \textbf{Acceptance}\\
        \midrule
        Total  & 13695 & 100.00\% \\
        + reconstructed pulse          & 12357 & 90.23\% \\
        + coincidence selection        & 9495  & 76.84\%  \\
        + anti-coincidence veto   & 6631  & 69.83\%  \\
        + width selection             & 1718  & 25.90\%  \\
        \bottomrule
    \end{tabular}
\end{table}

\section{Results}
\label{sec:results}

In this work, we demonstrate the drift of ionization electrons in LXe over a length of \SI{2.6}{\m} and measure the corresponding drift velocities at two different drift fields. The drift time is defined as the time elapsed between the muon trigger and the start of the signal peak detected by the SiPM array (see App.~\ref{app:reconstruction} for details of the peak-reconstruction procedure). For muons crossing the TPC, we observe two features at known depths in the drift time distribution. The first is a peak from ionization along the tracks selected by the coincidence. The second is a step at the largest drift times, from ionization along the full length of the TPC (horizontal and vertical muon showers), which marks the end of the TPC.

\begin{figure}[htbp]
    \centering
    \includegraphics[width=\columnwidth]{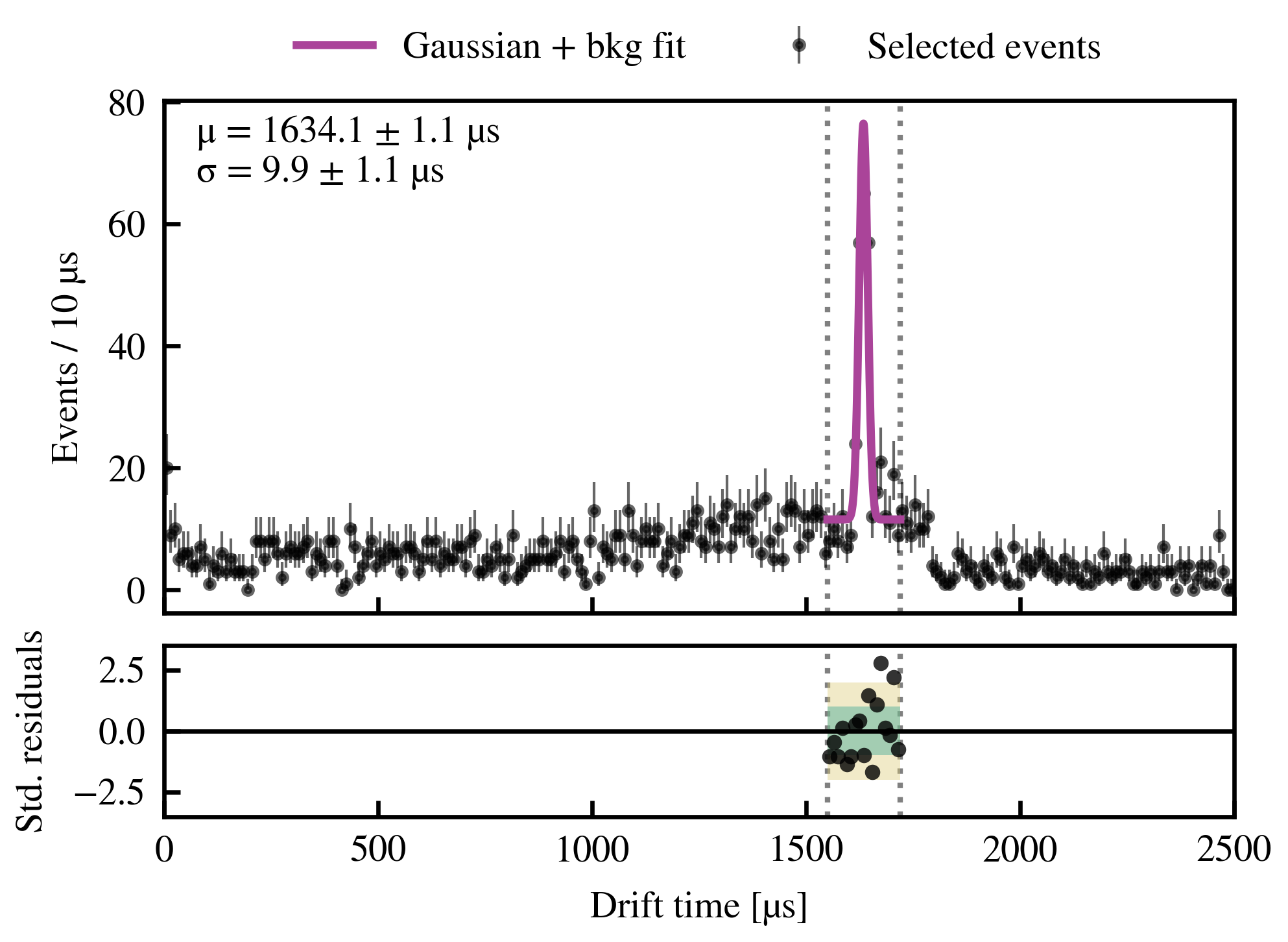}
    \caption{Distribution of drift times for the selected muon-coincidence events, in bins of \SI{10}{\micro\second}. The peak is fitted with a Gaussian plus a constant background. The lower panel shows the residuals for the bins in the fit range; the green and yellow bands indicate the $\pm1\sigma$ and $\pm2\sigma$ intervals respectively.}
    \label{fig:drift_hist}
\end{figure}

Figure~\ref{fig:drift_hist} shows the pulse drift time distribution of the muon coincidence events, which has a clear peak above a background. Fitting the peak yields the drift time of the electrons produced along the muon track, as the geometry of the coincidence selects tracks with a drift length of \SI{233.95 \pm 1.76}{\centi\meter}, as described in Sect.~\ref{sec:mnt_des}. The peak is fitted with a Gaussian plus a constant background, yielding $\mu = \SI{1634.1 \pm 1.1}{\micro\second}$ and $\sigma = \SI{9.9 \pm 1.1}{\micro\second}$  ($\chi^2/\mathrm{ndf} = 25.8/13$). The elevated $\chi^2$ is mainly driven by a few bins on the right-hand side of the peak that lie above the fitted background. However, the extracted parameters are robust against this modeling choice: they remain stable when varying the fit range or replacing the Gaussian with an exponentially tailed Gaussian. 
 
\begin{figure}[htbp]
    \centering
    \includegraphics[width=\columnwidth]{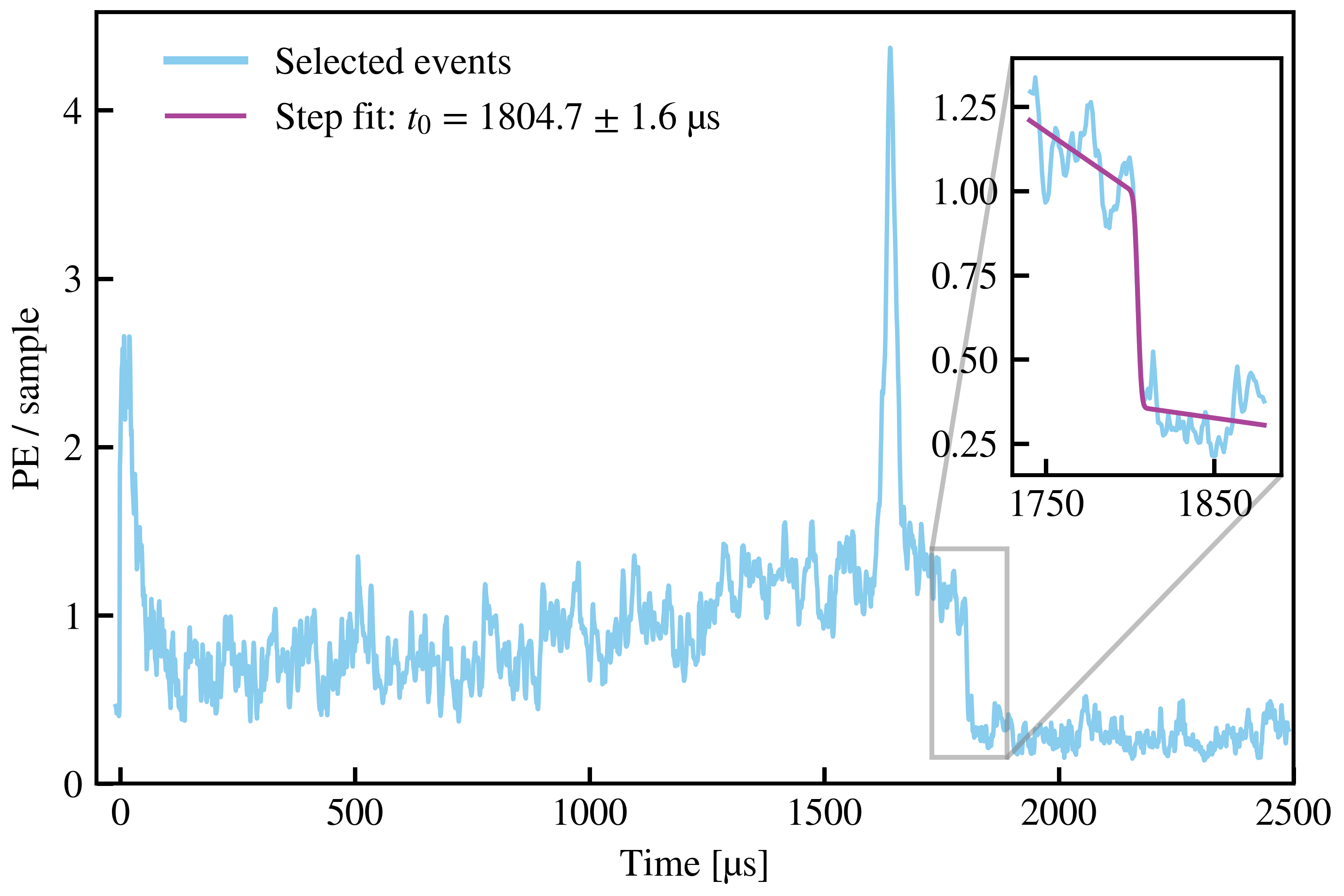}
    \caption{Average summed waveform of the selected muon-coincidence events. The inset shows the drop in the waveform, fitted with an error function, and yielding $t_0 = \SI{1804.7 \pm 1.6}{\micro\second}$. The peak around \SI{1600}{\micro\second}, corresponding to the muon-coincidence peak fitted in Fig.~\ref{fig:drift_hist}, is also clearly visible.}
    \label{fig:avg_wf}
\end{figure}

The drift time distribution shows a discontinuity after the main peak, which we attribute to the TPC boundary, i.e.\ to the full drift length. 
However, the drift time associated with this second feature would be underestimated if determined from the distribution of pulse start times. A more accurate estimate can be obtained from the average waveform of the selected events, shown in Fig.~\ref{fig:avg_wf}. A muon crossing the TPC produces ionization along its track and also emits light that can ionize particles across the column. The resulting electrons reach the readout at different times after the trigger, up to the maximum drift time corresponding to the full TPC length. In the average waveform, this appears as a sharp drop in the signal. We fit the drop with an error function and obtain $t_0 = \SI{1804.7 \pm 1.6}{\micro\second}$. The uncertainty combines the spread from resampling the data with the variation of the result for different fit windows, binning and fit models.

We determine the drift velocity from a $\chi^2$-fit to the muon drift time peak and TPC edge points. The uncertainty of the total TPC length due to liquid level variations, \SI{+-0.11}{\cm}, is accounted for by allowing the linear drift time-drift length relationship to include a non-zero constant term constrained by this value. The total uncertainty of the drift length of the fitted muon peak is calculated by adding the individual contributions in quadrature. These are the muon track variation (\SI{+-1.76}{\cm}/$\sqrt{N}$, where $N$ is the number of pulses) and the muon panel position measurement uncertainty (\SI{+-0.50}{\cm}). Drift time uncertainties for both the muon drift time peak and the TPC edge are taken from the respective drift time fits. The fit yields a drift velocity of 
\SI{1.436 \pm 0.001}{\milli\meter\per\micro\second} at an electric field of \SI{96.0 \pm 0.42}{\volt\per\centi\meter} and a liquid temperature of \SI{178.95 \pm 0.37}{\kelvin}. 

Additionally, two muon coincidence measurements at different heights from a previous run in 2025 performed at a liquid temperature of \SI{179.75\pm0.30}{\kelvin} and an electric field of \SI{80.92\pm0.36}{\volt\per\centi\meter} are included. Analogous to the 2026 run, the drift time is extracted by fitting the drift time distribution. The run conditions, datasets and resulting drift time distributions are described in the App.~\ref{app:data_2025}. A simultaneous fit to the two muon peaks and the liquid level as the zero point results in a drift velocity of \SI{1.320 \pm 0.006}{\milli\meter\per\micro\second}. The uncertainties are evaluated using the same procedure, with the total uncertainty of the muon drift lengths based on the muon track variation (\SI{+-2.57}{\cm}/$\sqrt{N}$) for the higher position and (\SI{+-2.51}{\cm}/$\sqrt{N}$) for the lower position and the panel position measurement uncertainty (\SI{+-0.50}{\cm}). The uncertainty of the zero point is (\SI{+-0.03}{\cm}), estimated from the variation of the liquid level. The lower statistics in the 2025 run result in a larger uncertainty on the measured drift velocity.

\begin{figure}
    \centering
    \includegraphics[width=1\linewidth]{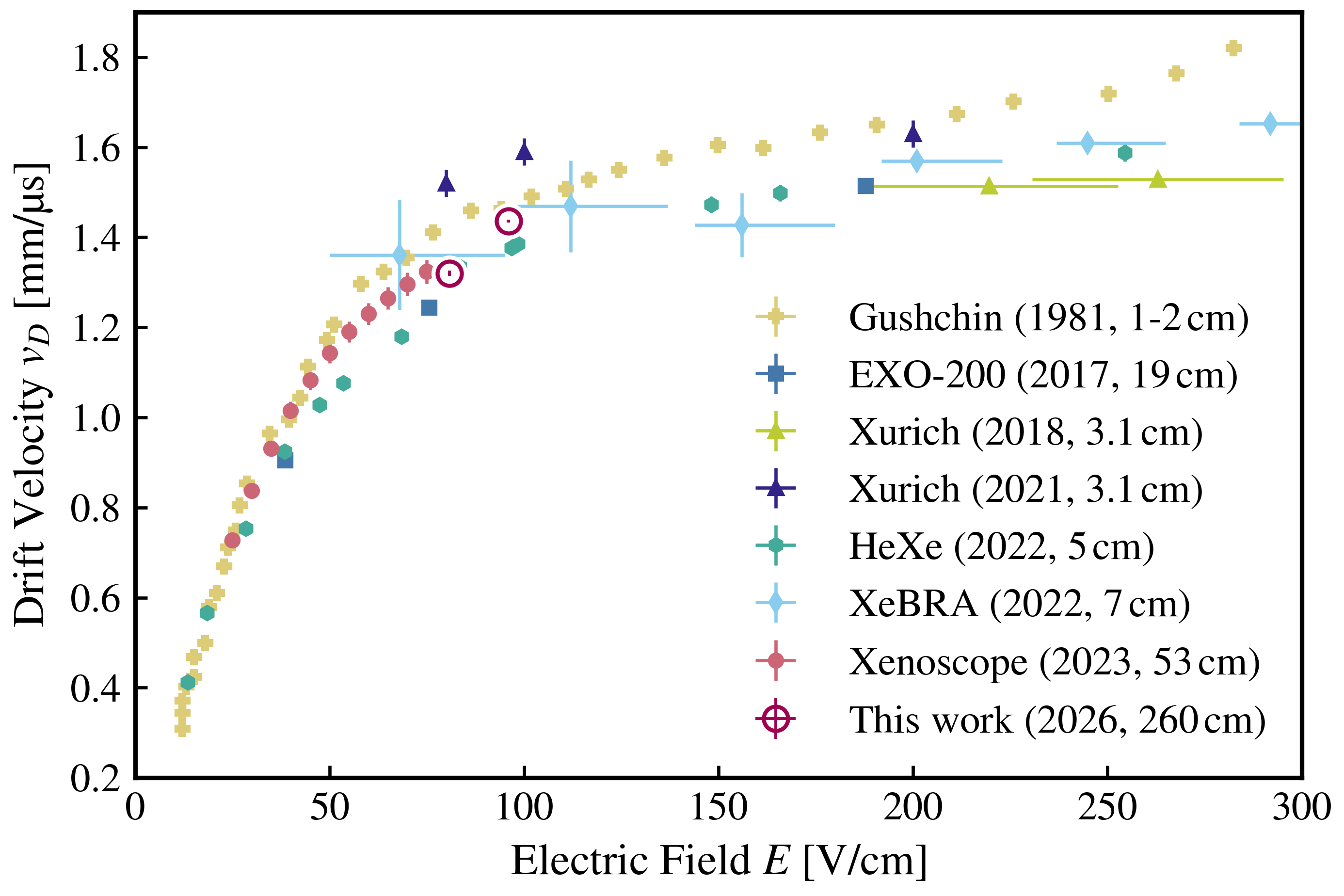}
    \caption{Drift velocity as a function of the electric drift field. The result is compared to the literature values from Gushchin (165K)~\cite{gushchin1981electron}, EXO-200 (167K)~\cite{PhysRevC.95.025502}, Xurich (2018, 184K)~\cite{Xurich2018}, Xurich (2021, 177K)~\cite{Thieme:2022dze}, HeXe~(174 K) \cite{Jorg:2021hzu}, XeBRA (173K)~\cite{Baur:2022sel} and Xenoscope (178K)~\cite{Baudis_2023}. The second number after the year is the drift length of each measurement.}
    \label{fig:DriftVelocity}
\end{figure}

Figure~\ref{fig:DriftVelocity} shows the measured drift velocities at the two drift fields, which are in good agreement with previous measurements reported in the literature. This result marks the first drift velocity calculation using electrons drifting further than \SI{1.5}{\m}.
\section{Conclusions and outlook}
\label{sec:conc}

We have observed ionization signals consistent with electron transport over the full \SI{2.6}{\meter} drift length of the Xenoscope TPC. Cosmic-muon coincidence measurements and the endpoint of the average S2 waveform provide complementary timing information. We obtain electron drift velocities of \SI{1.320 \pm 0.006}{\milli\meter\per\micro\second} at \SI{80.92\pm0.36}{\volt\per\centi\meter} and \SI{1.436 \pm 0.001}{\milli\meter\per\micro\second} at \SI{96.0 \pm 0.42}{\volt\per\centi\meter}. These measurements demonstrate detectable charge transport over distances relevant to next-generation liquid-xenon TPCs. Future measurements of electron attenuation, diffusion and optical transport of the VUV light will further characterize performance over these distances.


\begin{acknowledgements}
We gratefully acknowledge support from the Swiss National Science Foundation via the grants 20FL20-232696, 2000-1-243007 and 200021-228326. This work was also supported by the European Research Council (ERC) under the European Union’s Horizon 2020 research and innovation programme, grant agreement No. 742789 (Xenoscope). We thank the University of Zurich for continued support, as well as the mechanical and electronics workshop of the Department of Physics.  

\end{acknowledgements}

\section{Appendix}
\label{sec:appendix}

\subsection{Data taking 2025}
\label{app:data_2025}
In 2025 muon coincidence measurements analogous to this run were performed at two different heights and a drift field of \SI{80.92 \pm 0.36}{\volt\per\centi\meter}. Throughout the campaign, the LXe level was maintained at \SI{8.7}{\mm} above the gate electrode. The detector was operated at a temperature of \SI{179.75\pm0.30}{\kelvin}, a gas pressure of \SI{2.164\pm0.011}{bar}, and a recirculation flow rate of \SI{50}{\slpm}.

The SiPM array was operated at a bias voltage of \SI{48}{\V}. Tile F was not responding and therefore turned off for the entire run. We performed daily LED calibrations at different light levels to monitor the gain of each SiPM. Due to an uneven light distribution in Tile C and G and large fluctuations in the gain calculation resulting from limited statistics in the single-photoelectron peak, the average gain value over the full run is used. The gain values of the other tiles vary up to 2\%  making this an appropriate choice.

The muon panels were mounted at a \SI{~13}{\degree} angle to the horizon at two different positions. An overview of the different drift lengths and precise angles is given in Table \ref{tab:Run5_muonpanels}. Under the assumption of a drift velocity of \SI{1.28}{\milli\metre\per\micro\second}~\cite{Jorg:2021hzu} the vertical track length can be transformed to the intrinsic width of the S2. The drift lengths range from \SIrange{78.2}{87.1}{\cm} and \SIrange{128.2}{136.9}{\cm} for the two measurement positions, which we model using mean and standard deviations of \SI{82.5 \pm 2.57}{\centi\meter} and \SI{132.5 \pm 2.51}{\centi\meter}, respectively.

\begin{table}[htbp]
    \centering
    \caption{Drift length, coincidence angle, exposure and acceptance of each selection step of the muon-coincidence data.}
    \label{tab:Run5_muonpanels}
    \begin{tabular}{l r r}
        \toprule
        \textbf{Parameter} & \textbf{Measurement 1} & \textbf{Measurement 2} \\ 
        \midrule
        Drift length [\unit{\cm}]       & $82.5\pm2.57$ & $132.5\pm2.51$ \\
        Angle [\unit{\degree}]          & 13.0     & 12.2 \\
        Exposure [\unit{\hour}] & 38.74 & 44.36\\
        Total Number of Events & 466 & 631 \\
        + reconstructed pulse & 348 (74.7\%) & 521 (82.6\%)\\
        + coincidence selection & 318 (68.2\%) & 473 (75.0\%)\\
        + width selection & 128 (27.5\%) & 137 (21.7\%)\\
        \bottomrule
    \end{tabular}
\end{table}

The muon panels were operated with a bias voltage of \SI{1}{\kV} and the anti-coincidence veto was not available. The muon coincidence selection is tuned on the analysis dataset resulting in the condition that in case of a coincidence within \SI{500}{\nano\second}, the summed area of the muon panels has to be above 400\,ADC. The width based pulse selection is identical to the one explained in Sect.~\ref{sec:analysis}. The total number of measured events and the fraction of events passing each selection step is summarized in Table \ref{tab:Run5_muonpanels}.

\begin{figure}[htbp]
    \centering
    \begin{subfigure}[b]{\linewidth}
        \centering
        \includegraphics[width=\linewidth]{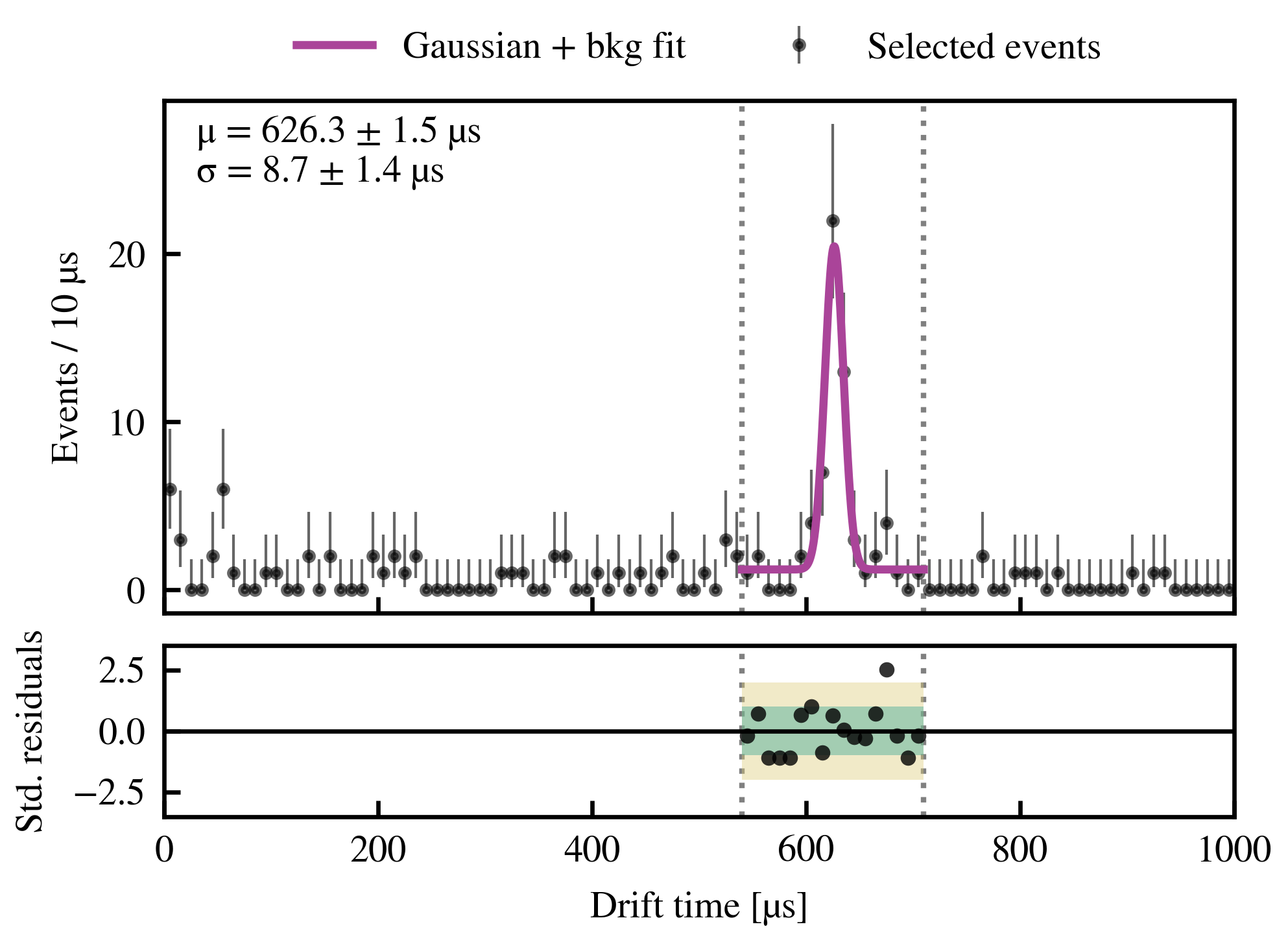}
        \caption{Distribution of the pulse drift time for \SI{82.5\pm2.57}{\centi\meter} drift.}
        \label{fig:Run5Pos2Dthist}
    \end{subfigure}
    
    \vspace{0.5cm}
    
    \begin{subfigure}[b]{\linewidth}
        \centering
        \includegraphics[width=\linewidth]{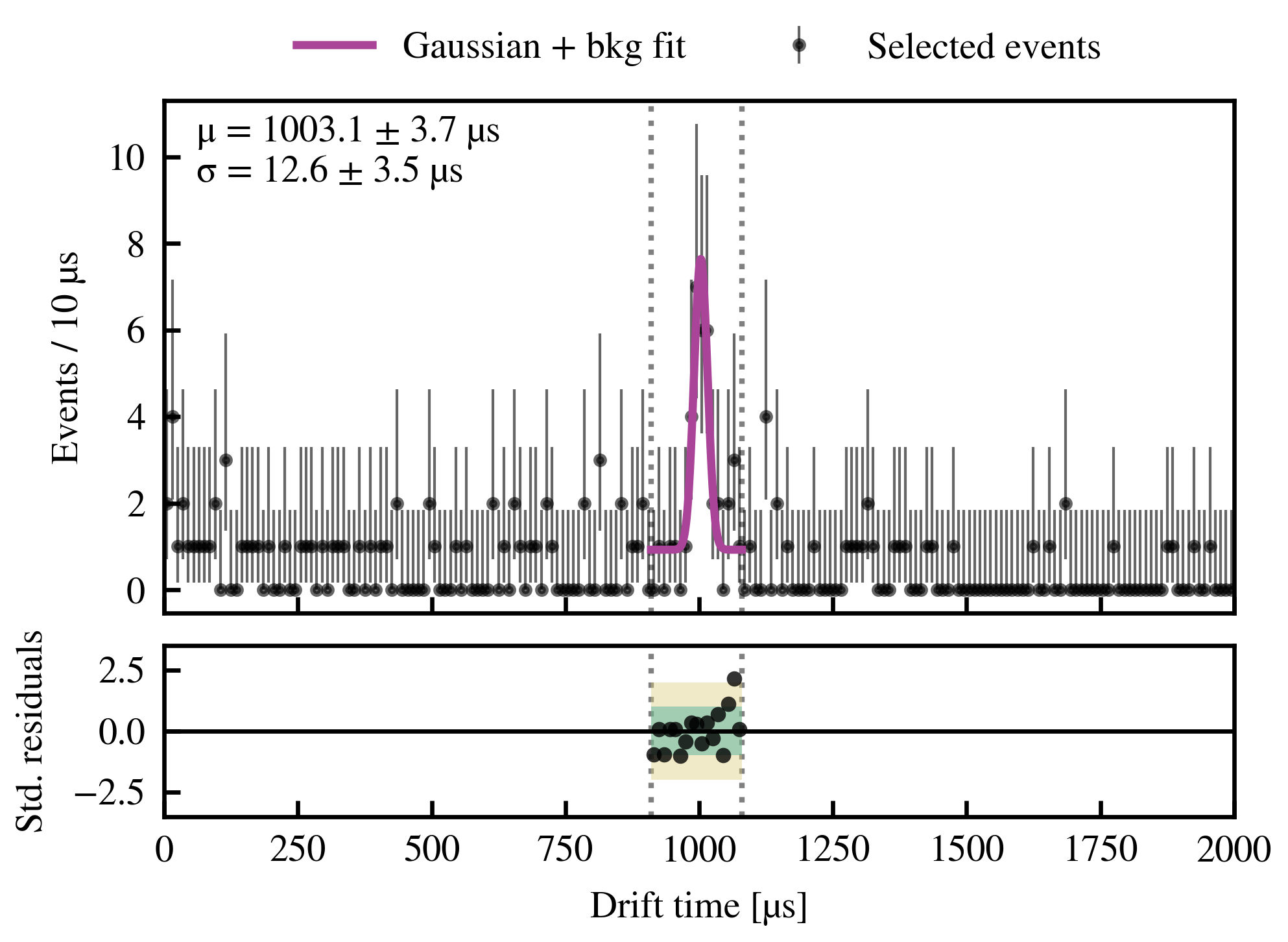}
        \caption{Distribution of the pulse drift time for \SI{132.5\pm2.51}{\centi\meter} drift.}
        \label{fig:Run5Pos3DriftTime}
    \end{subfigure}
    
    \caption{Pulse drift time distributions at different drift lengths. The peak is fitted with a Gaussian plus a constant background. The lower panel shows the residuals for the bins in the fit range; the green and yellow bands indicate the $\pm1\sigma$ and $\pm2\sigma$ intervals respectively.}
    \label{fig:Run5DriftTimes}
\end{figure}

The analysis dataset includes \SI{38.74}{\hour} for the lower position and \SI{44.36}{\hour} for the higher position. The resulting distribution of the pulse drift time after applying the trigger and width cut is plotted in Fig. \ref{fig:Run5DriftTimes}. In both positions there is a visible peak at different drift times. The average pulse drift time is extracted by fitting a Gaussian plus a constant background to each peak. These values together with the drift length are used to calculate the drift velocity explained in Sect. \ref{sec:results}.

\subsection{SiPM array characterization measurement}
\label{app:sipm}

A SiPM array characterization measurement was performed in the beginning of the run. Both the bias and LED voltages are scanned, using initial ranges of 48-53~V and 2.6-3.4~V, respectively, based on values established in previous Xenoscope runs. The bias-voltage scan identifies a common operating voltage suitable for all tiles, which share a single power supply. For the LED calibration, two optical fibers are used to illuminate the SiPM array. As the fibers are immersed in LXe, their positions may shift slightly, resulting in different amounts of light reaching the individual tiles. The LED-voltage scan therefore determines an illumination sequence that provides suitable signals across all tiles despite these differences. 

Since the SiPM gain scales linearly with the overvoltage, defined as the difference between the applied bias voltage and the breakdown voltage of each tile \cite{Peres_2023}, the latter can be determined from a linear fit to the gain as a function of bias voltage. The extracted breakdown voltages range from \SI{45.6}{\volt} to \SI{47.4}{\volt} across the twelve tiles. A common operating bias voltage of \SI{51.0}{\volt} therefore provides sufficient overvoltage for all tiles. 

\begin{figure}
    \centering
    \includegraphics[width=\linewidth]{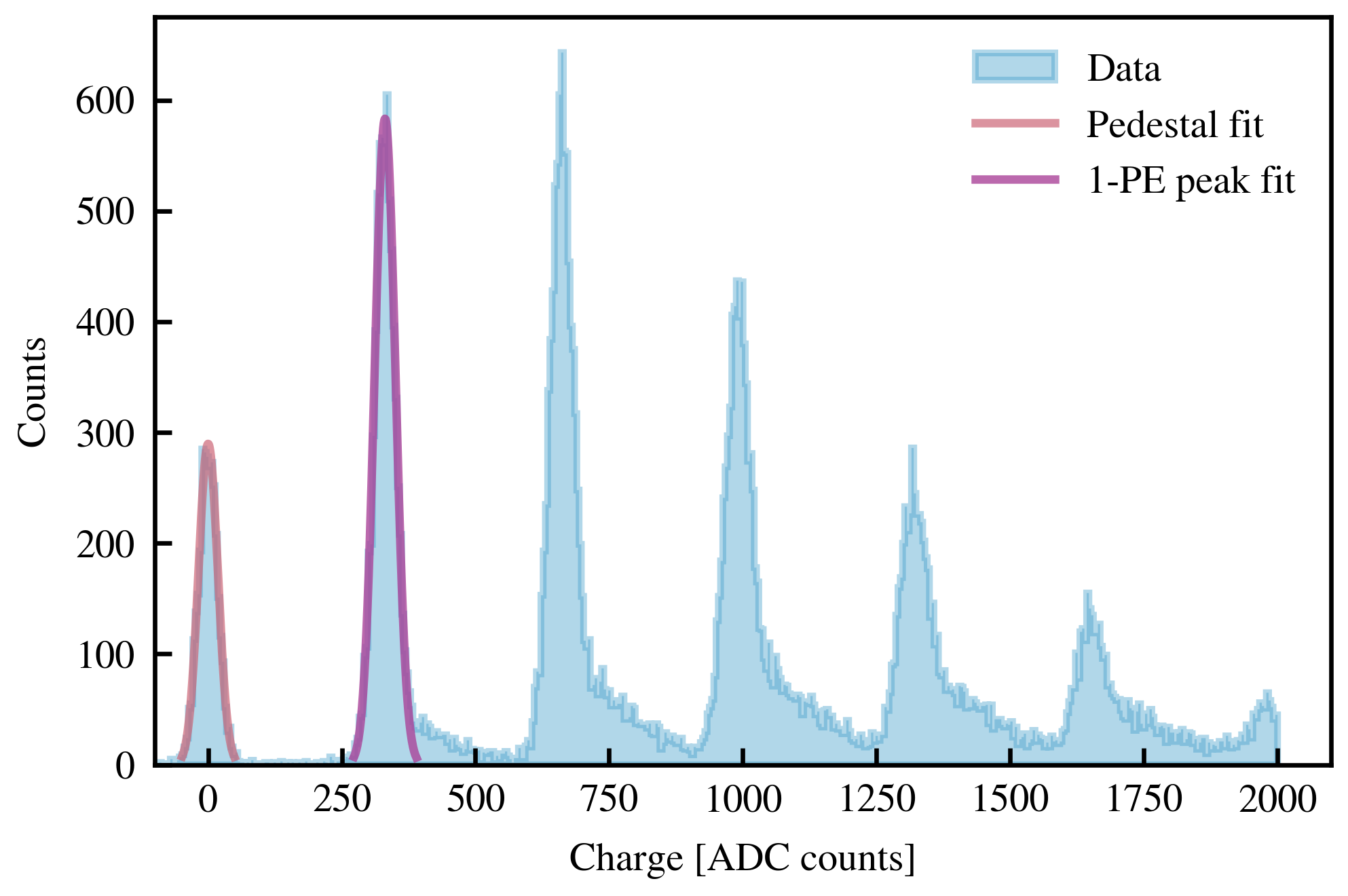}
    \caption{Charge spectrum of tile E at a bias voltage of \SI{51.0}{\volt} and LED voltage of \SI{2.95}{\volt}, taken during the SiPM tile characterization campaign. The pedestal and SPE peaks are shown with their respective Gaussian fits.}
    \label{fig:pe_spectrum_example}
\end{figure}

When the bias voltage is set, the LED voltage is varied to identify suitable light levels. Fig.~\ref{fig:pe_spectrum_example} shows an example of a satisfactory SPE spectrum, where the individual PE peaks are clearly resolved. The resulting parameters are then implemented in the gain calibration sequence of the DAQ. During a run, the user can activate this sequence to acquire LED calibration data, which are stored on disk and subsequently processed offline to extract the SiPM gains.

The acquired waveforms are processed using a custom analysis framework. For each tile, a charge spectrum is constructed from the corresponding waveforms by subtracting the baseline, determined from a pre-signal region, and integrating the waveform over the signal region. The positions of the pedestal and the SPE peak are then determined from each spectrum, and their separation provides the charge per PE. This charge is converted to the SiPM gain using the known characteristics of the readout electronics, including the ADC dynamic range and resolution, the input impedance, and the total amplification of the signal chain. 

The resulting gain values are shown in Fig.~\ref{fig:sipm_stability}. The twelve tiles exhibit different absolute gain values, as also observed during the commissioning run, ranging from $1.2\times10^6$ to $1.6\times10^6$. The gain of each individual tile remains stable throughout the run, with a relative standard deviation below 0.7$\%$ for all tiles. Most tiles show an SPE resolution between 6$\%$ and 8$\%$, with tile M standing out as an outlier at 11$\%$. 

\subsection{Peak reconstruction}
\label{app:reconstruction}

\begin{figure}
    \centering
    \includegraphics[width=\linewidth]{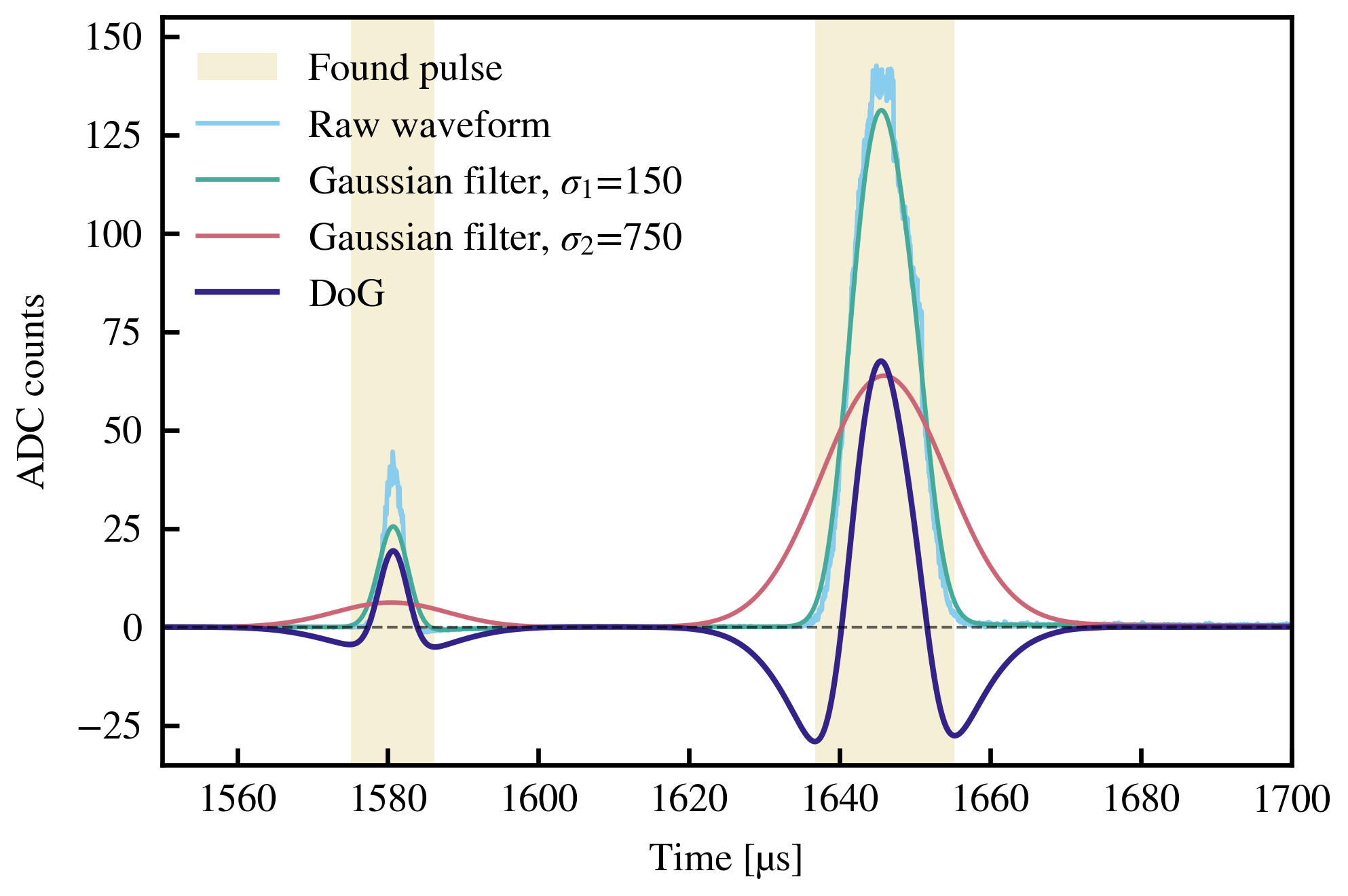}
    \caption{Example of the DoG filtering procedure applied to the summed top-array waveform of a single-S2-like event, with the nominal smoothing scales.}
    \label{fig:dog_example}
\end{figure}

As the first step of processing towards obtaining clean events, we develop a pulse-finding algorithm and merge pulses into peaks.
Pulses are identified in the summed top-array waveform using a DoG filter: the raw waveform is smoothed with two Gaussian kernels of different widths ($\sigma_1 < \sigma_2$), and the narrower smoothing is subtracted from the wider one, see example in Fig.~\ref{fig:dog_example}. The DoG is a feature-finding method equivalent to a bandpass filter, whose sensitivity to features is set by the two smoothing scales $\sigma_1$ and $\sigma_2$, the main parameters of the algorithm. Smaller values make the finder more sensitive to fine structure, but also to noise, while larger values smooth over noise but resolve fewer features.
In the context of this study, S2-like peaks at different drift times are inspected, but we do not attempt to pair S1 and S2 into events, as S1 pulses are occasionally visible in the raw waveforms but detected at a much lower rate due to the limited efficiency of the top array in collecting such faint light signals. Candidate pulse positions and boundaries are then found from the derivatives of the filtered waveform.

\begin{figure}[htbp]
    \centering
    \includegraphics[width=\linewidth]{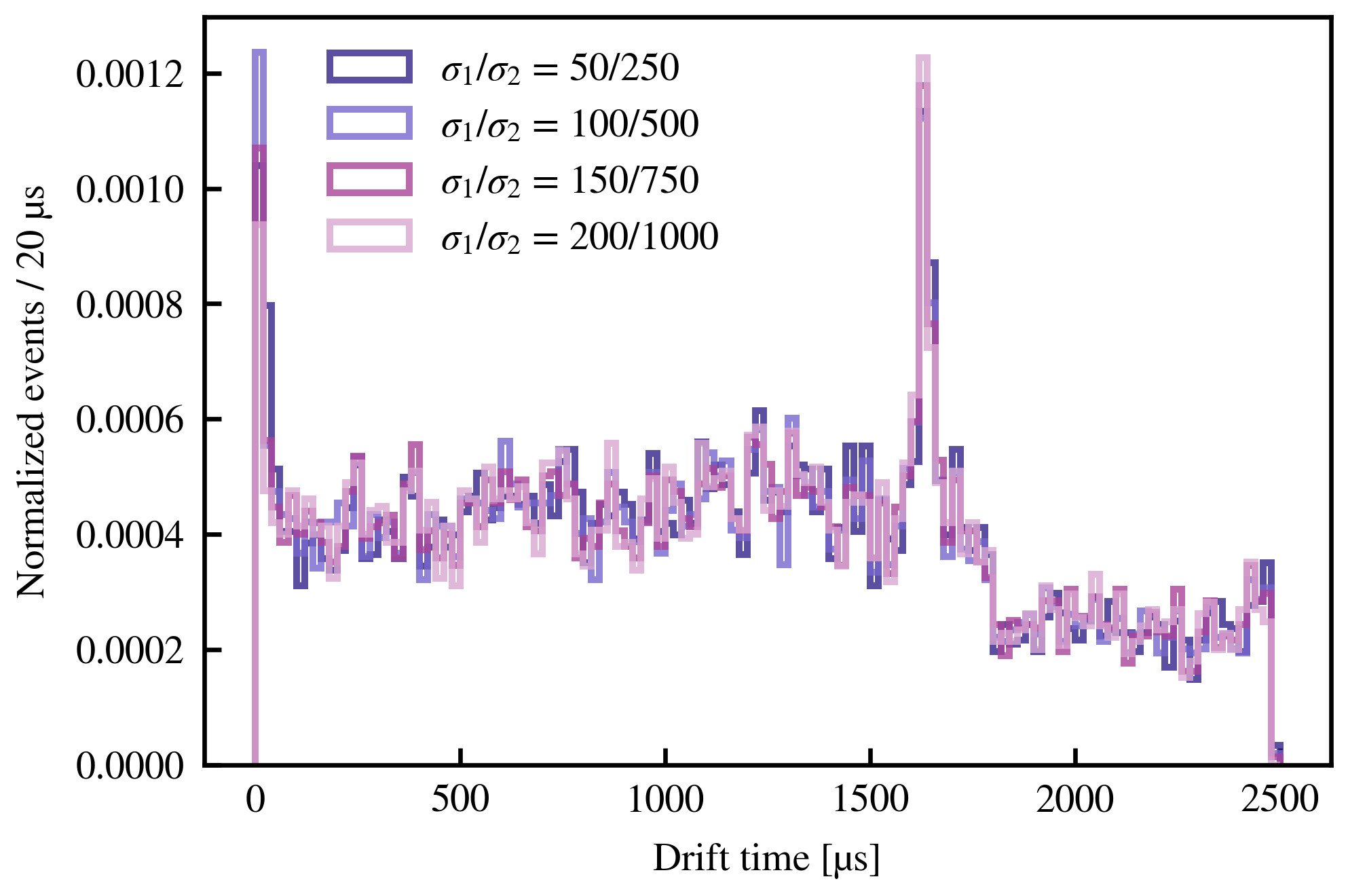}
    \caption{Drift time distribution of all pulses found in the triggered datasets, for different choices of the smoothing scales $(\sigma_1, \sigma_2)$, showing that the results are not strongly sensitive to the exact choice of smoothing parameters.}
    \label{fig:pulse_finding_robstness}
\end{figure}

We inspect the pulse-finding procedure on individual waveforms with the four different ratios shown in Fig.~\ref{fig:pulse_finding_robstness}, with sensitivity to temporal width---computed from the peak sensitivity response of the filter---ranging from \SI{1.4}{\micro\second} to \SI{5.5}{\micro\second}.
Since the expected S2 width is \SI{\sim 27}{\micro\second}, and given the observed variability in S2 substructure, the filter must be able to resolve both fine sub-structure within a single S2 when present, while also correctly identifying S2s composed of a single peak.
The nominal parameters were chosen to preserve sensitivity to smaller pulses while suppressing the bulk of the noise, corresponding to $(\sigma_1, \sigma_2) = (150, 750)$ samples. For these parameters, the peak sensitivity of the filter corresponds to a timescale of around  \SI{4.1}{\micro\second}. 
As shown in Fig.~\ref{fig:pulse_finding_robstness}, the results are robust against variations of these parameters: the number of events with a reconstructed drift time changes by less than 2\%, and the position of the drift-time peak is unchanged.

In the pulse-finding algorithm, each identified pulse is assigned a start and end boundary at the minima of the DoG response, defining a candidate. These candidates are subsequently merged into a single S2 peak if the gap between the end boundary of one pulse and the start boundary of the next falls within a threshold of 100 samples, i.e. \SI{1}{\micro\second}. 
The merged S2 is in turn assigned a start and end boundary, and its start is used in the definition of the drift time used in the analysis. 

\subsection{Muon trigger tuning}
\label{app:muon_coin_tuning}

Following the run, we conducted a muon panel characterization campaign. Both panels had their configuration unchanged while measurements were taken with OR and AND trigger logic. The measurements are used to tune a secondary trigger selection applied in the analysis and an anti-coincidence veto to target vertical muon showers.

\begin{figure}
    \centering
    \includegraphics[width=1\linewidth]{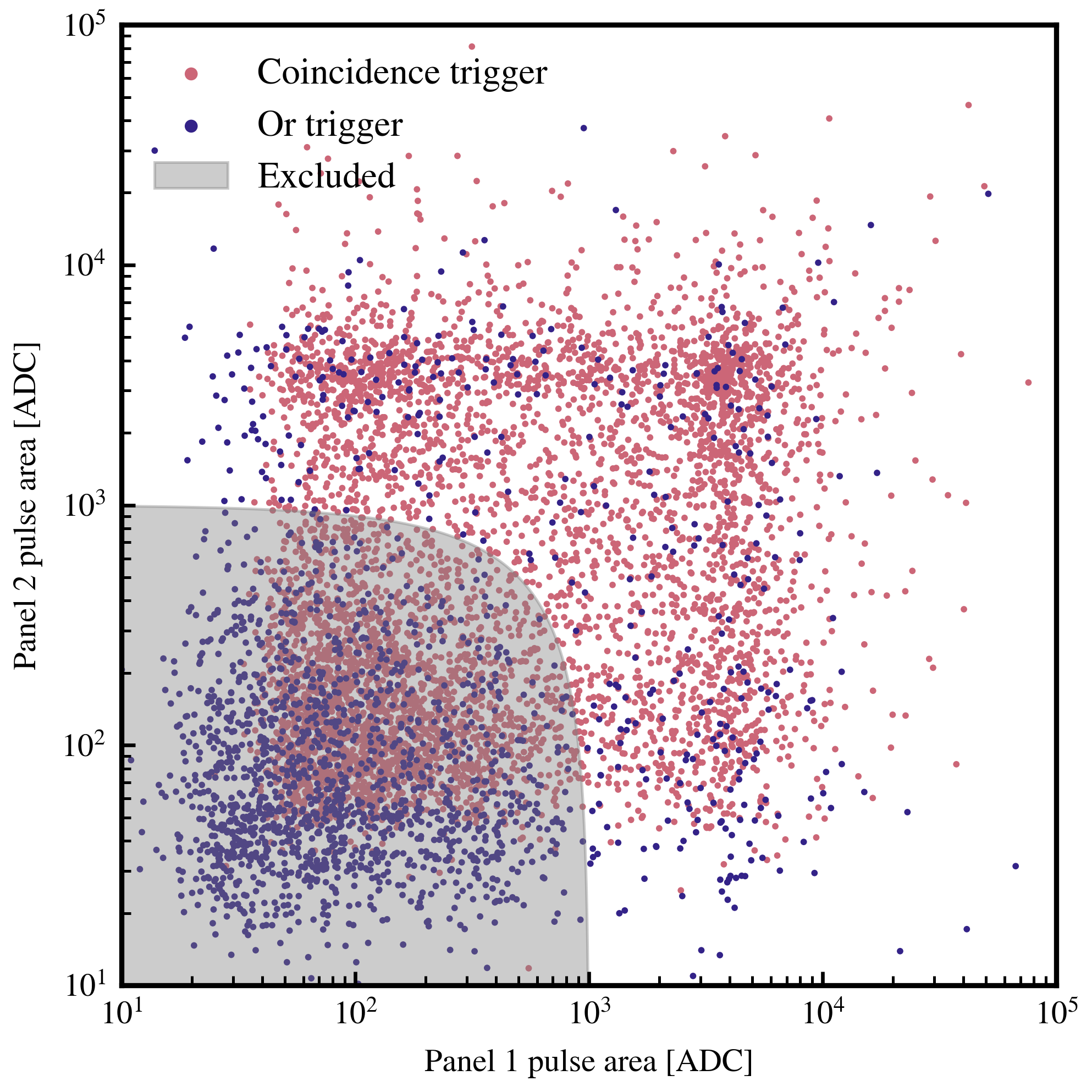}
    \caption{Event selection based on the pulse area in the coincidence muon panels targeting the accidental gamma ray coincidence background. The signal efficiency is 56.41\% and the background efficiency is 3.01\%. }
    \label{fig:MuonTriggerTuning}
\end{figure}

Each waveform is baseline-corrected using its pre-trigger window. A threshold-based pulse finder is then applied within a \SI{500}{\nano\second} window to identify the trigger pulse and integrate its area. The area distribution in the two coincidence muon panels is shown in  Fig. \ref{fig:MuonTriggerTuning}.  The OR trigger dataset in purple consists of a large population in the low area region, which is mainly originating from gamma rays. In the AND (coincidence) trigger condition the gamma ray population is still present with an additional population in the higher area region consisting of muons. The horizontal and vertical leakage into the low area region can either be from grazing muons creating only small pulses in one of the scintillators or an accidental coincidence of a muon and a gamma ray. In order to keep as much signal as possible the selection is loosely set to remove events where the summed area in both muon panels is below 1000\,ADC within the coincidence window of \SI{500}{\nano\second}. The efficiency for the OR triggered events is 3.01\% while 56.41\% of the AND triggered events are kept. 

Similarly the anti-coincidence veto is tuned based on the area in the anti-coincidence panel above the TPC. Events with pulse areas above 140\,ADC within the \SI{500}{\nano\second} coincidence window indicate a muon crossing the TPC vertically and are removed. 

\subsection{Drift field simulation}

The dual-phase design of the Xenoscope TPC relies on two distinct field regions to convert an interaction into a detectable signal. Electrons produced at the interaction site first drift upward through the bulk liquid, over the full length of the TPC, before reaching the liquid surface. There, a stronger extraction field pulls them across the liquid-gas interface into the gas phase, where they induce electroluminescence: a secondary scintillation signal collected by the photosensor array.
These two regions are defined by three electrodes: a cathode at negative potential at the bottom of the TPC and a gate \SI{\sim 260}{\centi\meter} above it, which together define the drift field, and an anode \SI{12}{\milli\meter} above the gate, held at positive potential to establish the extraction field. The electric field in the drift region is homogenized using 173 equidistant copper field-shaping rings. These are connected between the cathode and gate electrodes via a resistor chain, such that they follow the linear potential drop between the two electrodes.

While this configuration is designed to produce a uniform drift field over the bulk of the TPC, deviations can arise from the electrode geometry, the field-shaping rings structure, and the proximity of the strong extraction field to the drift region near the gate. This motivates the use of simulation tools to assess potential field non-uniformities and their impact on the calculated drift velocity. 

\begin{figure}
    \centering
    \includegraphics[width=0.6\linewidth]{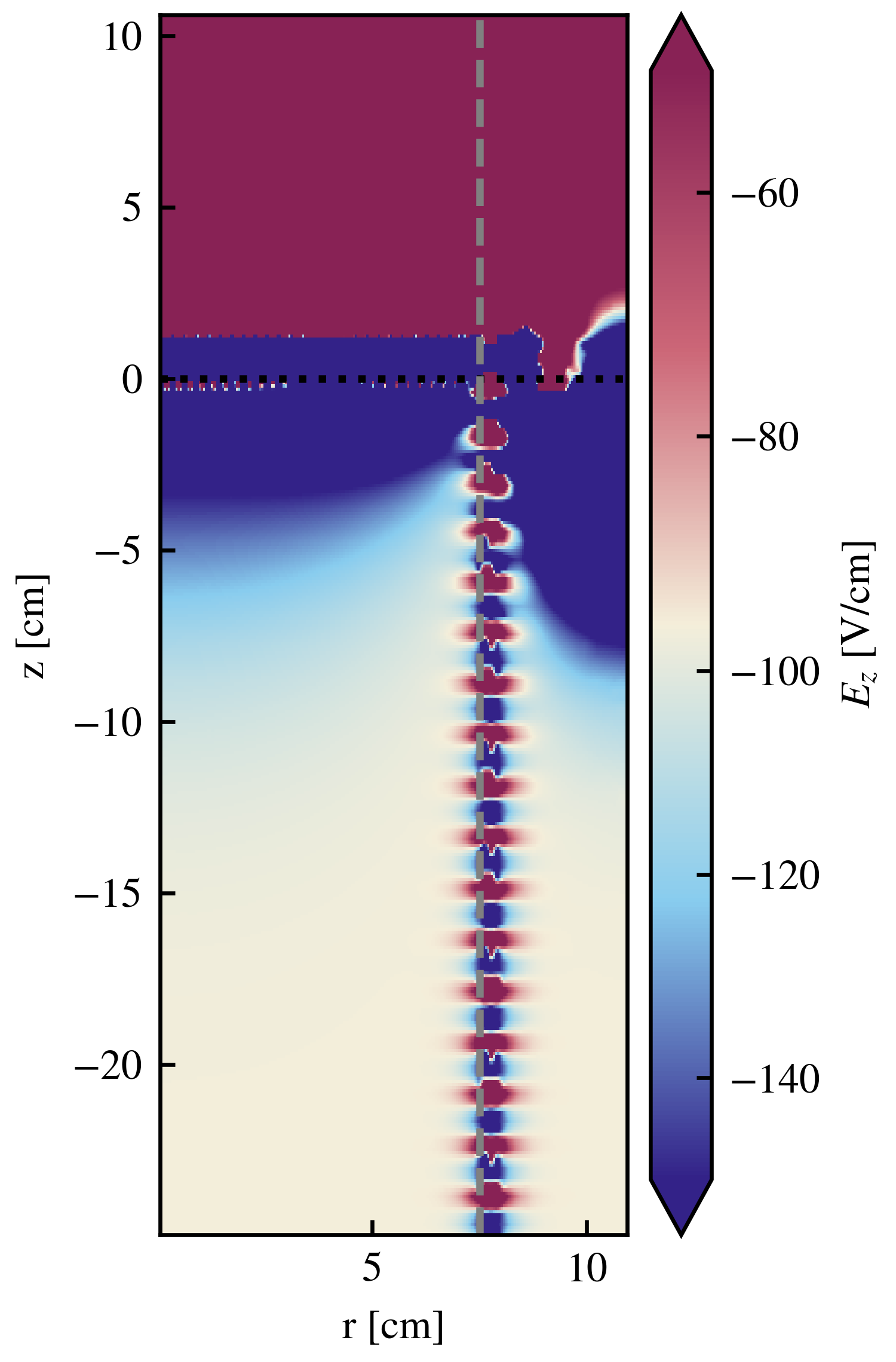}
    \caption{Axial component of the electric field, $E_z$, in the region around the gate, as obtained from the COMSOL simulation (with $z=0$ at the gate). The drift field of \SI{95.6 \pm 1.1}{\volt\per\centi\meter} is recovered in the bulk, with the field rising towards the gate, while the saturated region at the top corresponds to the extraction field between the gate and the anode. The periodic structure along the right-hand side comes from the field-shaping rings.}
    \label{fig:Ez_map}
\end{figure}

A model of the TPC, including all relevant components, is implemented and simulated using the commercial finite element software COMSOL Multiphysics \cite{COMSOL} (version 6.4). The simulation parameters are set to the values used during the science run, with the liquid level set to its average value of \SI{9.7}{\milli\meter} over the duration of the run, and the electrode potentials set to \SI{-29}{\kilo\volt} for the cathode, \SI{-4.2}{\kilo\volt} for the gate, and \SI{4.2}{\kilo\volt} for the anode.
Given that the detector is approximately axially symmetric, a 2D axisymmetric model is used. The simulation is performed through a fine triangular mesh, with a finer resolution near the gate and anode, where larger electric-field variations are expected. The resulting axial electric-field distribution around the gate is shown in Fig.~\ref{fig:Ez_map}.

The simulation reveals two distinct regions in terms of the drift-field configuration: the bulk of the TPC and the region near the gate.
From the axial profile of $E_z$, we find that the field deviates from flat within \SI{\sim 15}{\centi\meter} of the gate, where the proximity to the strong extraction field distorts the drift field. A comparable deviation is also observed near the cathode. Therefore, we define the bulk drift region as \SI{-255}{\centi\meter} $\le z \le$ \SI{-15}{\centi\meter} (with $z=0$ at the gate), together with a radial cut $r \le$ \SI{7}{\centi\meter}, \SI{0.5}{\centi\meter} inside the field-cage ring radius.
Over this volume, the average drift field is \SI{95.6 \pm 1.1}{\volt\per\centi\meter}, where the uncertainty is the point-to-point spatial variation of the field across the volume. 
Radially, the field is uniform to within $1\%$ out to an effective radius $R_\mathrm{eff} \approx \SI{6.4}{\centi\meter}$. For comparison, without excluding the near-electrode regions the average field over the full path up to the gate is \SI{97.0 \pm 9.3}{\volt\per\centi\meter}, with a spread of $9.6\%$, illustrating the extent to which the near-electrode distortions would degrade the uniformity. 

The drift velocity extracted from the measured drift time reflects the electric field along the full drift path, including the rapid increase near the gate. To account for this effect, we calculate the harmonic mean of $|E_z|$ along the drift path, averaged over radii $r \leq \SI{7}{\centi\meter}$, and we obtain a path-averaged field of \SI{96.4}{\volt\per\centi\meter}. We use this value together with the bulk field of \SI{95.6}{\volt\per\centi\meter} to determine the drift field reported in Fig.~\ref{fig:DriftVelocity}. Taking the midpoint and half the difference of these values gives $E=$ \SI{96.0 \pm 0.42}{\volt\per\centi\meter}. The two field estimates differ by less than $1\,\%$, reflecting the limited contribution of the near-gate region to the total drift path.

The same procedure applied to the data taking period in 2025 operating conditions yields a drift field value of $E=$ \SI{80.92 \pm 0.36}{\volt\per\centi\meter}.

\bibliographystyle{JHEP}
\bibliography{mybibfile}

\end{document}